\documentclass[preprint,amsmath,amssymb,prb]{revtex4}
\usepackage{stmaryrd}
\usepackage{amsmath}
\usepackage{amssymb}
\usepackage{graphicx}
\usepackage{dcolumn}
\usepackage{bm}
\usepackage{multirow}
\usepackage{hyperref}

\begin{document}

\begin{titlepage}

\title{Effect of Hartree-Fock Pseudopotential on the First-principles Electronic Structure}

\author{Hengxin Tan$^1$, Yuanchang Li$^2$\footnote{yuancli@bit.edu.cn}, S. B. Zhang$^{3,4}$\footnote{zhangs9@rpi.edu}, and Wenhui Duan$^{1,5}$\footnote{dwh@phys.tsinghua.edu.cn}}
\affiliation{$^1$State Key Laboratory of Low-Dimensional Quantum Physics and Collaborative Innovation Center of Quantum Matter, Department of Physics, Tsinghua University, Beijing 100084, China \\$^2$Advanced Research Institute of Multidisciplinary Science, Beijing Institute of Technology, Beijing 100081, China \\$^3$Beijing Computational Science Research Center, No.10 East Xibeiwang Road, Beijing 100193, China \\$^4$Department of Physics, Applied Physics and Astronomy, Rensselaer Polytechnic Institute, Troy, NY, 12180, USA \\$^5$Institute for Advanced Study, Tsinghua University, Beijing 100084, China}
\date{\today}

\begin{abstract}
Density functional theory (DFT) can run into serious difficulties with localized states in elements such as transition metals with occupied-$d$ states and oxygen. In contrast, Hartree-Fock (HF) method can be a better approach for such localized states. Here, we develop HF pseudopotentials to be used alongside with DFT for solids. The computation cost is on par with standard DFT. Calculations for a range of II-VI, III-V and group-IV semiconductors with diverse physical properties show observably improved band gap for systems containing $d$-electrons, whereby pointing to a new direction in electronic theory.

\vspace{8mm}
\end{abstract}

\maketitle
\draft
\vspace{2mm}

\end{titlepage}

\vspace{0.3cm}
\textbf{1. Introduction}
\vspace{0.3cm}

Density functional theory (DFT) has achieved great success in the electronic structure calculation \cite{PNAS2017} of solids by virtue of its commonly-accepted accuracy and efficiency. However, the exact form of the exchange-correlation functional is still unknown \cite{Medvedev}. As an approximation, usually the local density approximation (LDA) or the generalized gradient approximation (GGA) has been used. These approximations make the DFT a valuable tool, but also reveal its shortcoming in characterizing the material properties. A well-known example is the underestimation of band gap, especially for solids containing localized $d$-electrons. For instance, the DFT gap of wurtzite ZnO is only $\sim$0.9 eV, which is severely underestimated from the experimental value by $\sim$2.5 eV \cite{Madelung}. In the case of CdO with an indirect gap of 0.8 eV, DFT even yields qualitatively wrong result by predicting a semi-metal \cite{Burbano}. The underestimation of the band gap can severely undermine our ability to study defect physics and optical physics \cite{Toroker,SBZPRB63,Janotti}. To overcome these shortcomings, several methods \cite{Baraff,Anisimov,Aulbur,SBZJPCM,HSE} have been developed, among which the Heyd-Scuseria-Ernzerhof (HSE) hybrid functional \cite{HSE} approach has attracted much attention for its relatively accurate band gap and semicore $d$ states, by using a screened Coulomb potential for the Hartree-Fock (HF) exchange. In accordance, however, the computational cost is also significantly increased from that of DFT for the use of non-local functional for bulk materials. Often pseudopotentials (PPs) are used in DFT calculations, which reduces the number of electrons to be calculated whereby lowering the overall computational cost. Although the inclusion of the HF exchange has been shown to significantly improve the band gap \cite{HSE}, the HF is only applied to valence electrons but the PPs are still generated by standard LDA/GGA.

A common belief for the underestimated DFT band gap in $d$-electron systems is the unphysical self-interaction due to the local mean-field treatment. Despite also being a mean-field approach, the HF approximation does not suffer from such an error so it yields a larger band gap. This raises the question why not use HF PPs for elements for which the self-interaction error dominates over the correlation effect in their semicores. On the other hand, an all-electron hybrid functional calculation was reported very recently and it showed a small albeit consistent improvement in accuracy \cite{YangHY}. Given the remarkably distinct relaxation effect of core and valence electrons \cite{Chong}, including the same percentage of HF (25\%) as that of outer valence electrons is probably insufficient to treat the core electrons, responsible for the marginal change of the electronic structure. In this regard, it is instructive to explore an alternative hybrid functional calculation, i.e., the PP purely constructed from HF in combination with the DFT approximation for valence electrons, in order to deepen the understanding of HF exchange on the electronic structure. Such a treatment inherently has the advantage of computational efficiency as the cost is on par with standard DFT.

Note that, although since the birth of {\sl ab initio} PPs, we have been acquainted with the practice of generating the PPs using the same functional as the one used for solid-state calculations, such a consistency is not required. As a matter of fact, not only has this tradition been abandoned in hybrid functional calculations, but also the concept of optimized effective PP has been around for some time now \cite{kollmar2007}, which disconnects the issue of how to obtain PPs from the issue of how to apply them to condensed matter.

Following such a spirit, in this work, we develop HF PPs and apply them to electronic structure calculations of solids. Our approach may also be viewed as a hybrid approach, i.e., we use the HF method to generate the PPs, while using standard DFT for bulk study. However, there is an important difference from other hybrid functionals, namely, our approach is free of mixing parameters. The results on II-VI, III-V and group-IV semiconductors reveal a systematical improvement on the band structure, in particular, on the band gap. For examples, the band gap of wurtzite ZnO (zinc-blende ZnO) is increased from 0.86 eV to 2.13 eV (0.68 eV to 1.85 eV), while the Zn $3d$ states are pushed down to deep energies. The band gap of rocksalt CdO, on the other hand, is increased from 0 to 0.69 eV (versus 0.84 eV by experiment). By an intensive investigation of typical binary semiconductors, we provide a general guideline for the optimal choice of the PPs (between HF and PBE) over a wide range of elements. Moreover, the HF PP serves as a better starting potential for DFT+$U$ and HSE calculations, for instance, for ZnO it yields a remarkable band gap in agreement with experiment without having to adjust any parameters empirically. Finally, we stress that our hybrid scheme using HF PP alongside with standard DFT for solids maintains the cost efficiency of the DFT, and is thus particularly suited for large-scale systems such as defects and heterostructures.

\vspace{0.3cm}
\textbf{2. Methodology and models}
\vspace{0.3cm}

In our study, the HF approach, as implemented in the OPIUM code developed by Rappe group \cite{YangHY,Saidi,OPIUM}, was used to generate the PPs, where the scalar relativistic effect was included, but not the spin-orbit interaction (See the Supplemental Material for more details of the HF PPs.). For this reason, heavy elements such as Sn and Pb were not considered here. Unless specified, bulk calculations were carried out by using the Quantum ESPRESSO code \cite{QE} with the Perdew-Burke-Ernzerhof (PBE) \cite{PBE} exchange correlation functional. The II-VI, III-V and group-IV semiconductors typically have the diamond, zinc-blende, wurtzite, and rocksalt structures with a few exceptions. The experimental lattice constants were used. The energy cutoff was set to 60 Ry. The total energy convergence criteria was 10$^{-6}$ Ry/cell. The 10$\times$10$\times$10, 12$\times$12$\times$8 and at least 8$\times$8$\times$8 \emph{k}-meshes were respectively used for the diamond and zinc-blende structures, wurtzite structures, and the rest structures.

\begin{table}
\caption{Band gaps calculated by different PP combinations and the corresponding experimental values, in eV, as well as the band gap type, direct (D) or indirect (I). The systems are crystallized in diamond (DM), Zinc-blende (ZB), Wurtzite (WZ) and the other structures including rocksalt MgO and MnO, rutile (R) and anatase (A) TiO$_2$, and cubic bixbyite In$_2$O$_3$ ($^*$Note that CdO is crystalized in rocksalt structure where we put it in ZB just for comparison). In the ``Method" column, the optimal PP combination is indicated with the former PP for cation and the later for anion. The boldface systems are those with sizable band gap corrections by HF PPs relative to the results of PBE ones.}
%\renewcommand\arraystretch{1.4}
%\addtolength\tabcolsep{6pt}
\begin{tabular}{|c|c|c|c|c|c|c|c|c|c|c|}
%\hline
\hline
 \multirow{2}{*}{} & \multirow{2}{*}{System} & \multicolumn{2}{c|}{Band gaps} & \multicolumn{2}{c|}{Optimal} & \multirow{2}{*}{Exp.} & \multirow{2}{*}{~I/D~}\\
 \cline{3-6}
        & & ~HF PP~ & PBE PP & Method & ~~Gap~~ & & \\
 \hline
 \multirow{2}*{DM}& Si  & 0.63 & 0.59 & HF & 0.63 & 1.17$^a$ & I \\
        & Ge  & 0.004 & 0  & HF & 0 & 0.74$^a$ & I \\
 \hline
 \multirow{12}*{ZB}& GaN & 1.98 & 1.81 & HF+HF & 1.98 & 3.30$^{b}$ & D \\
        & GaP & 1.16 & 1.61 & PBE+PBE & 1.61 & 2.35$^a$ & I \\
        & {\bf GaAs}& 1.04 & 0.51 & HF+HF  & 1.04 & 1.52$^a$ & D \\
        & InN & 0    & 0    &  -     & -    & 0.78$^b$ & D \\
        & InP & 0.41 & 0.69 & HF+PBE & 0.99 & 1.42$^a$ & D \\
        & {\bf InAs}& 0.24 &  0   & HF+HF  & 0.24 & 0.42$^a$ & D \\
        & {\bf ZnO} & 1.85 & 0.68 & HF+HF  & 1.85 & 3.27$^c$ & D \\
        & ZnS & 2.40 & 2.10 & HF+PBE & 2.77 & 3.72$^a$ & D \\
        & ZnSe& 1.61 & 1.27 & HF+PBE & 1.91 & 2.82$^a$ & D \\
        & {\bf CdO$^*$}& 0.69 &  0   &  HF+HF & 0.69 & 0.84$^a$ & I \\
        & CdS & 1.20 & 1.15 & HF+PBE & 1.66 & 2.5$^d$ & D \\
        & CdSe& 0.84 & 0.63 & HF+PBE & 1.12 & 1.74$^a$ & D \\
 \hline
 \multirow{5}*{WZ} & GaN & 2.35 & 2.16 & HF+HF & 2.35 & 3.50$^a$ & D \\
        & {\bf InN} & 0.10 & 0.02 & HF+PBE & 0.11 & 0.78$^b$ & D \\
        & {\bf ZnO} & 2.13 & 0.86 & HF+HF  & 2.13 & 3.44$^a$ & D \\
        & ZnS & 2.49 & 2.17 & HF+PBE & 2.84 & 3.91$^a$ & D \\
        & CdS & 1.28 & 1.22 & HF+PBE & 1.74 & 2.48$^a$ & D \\
 \hline
 \multirow{5}*{Others} & MgO & 5.13 & 4.71 & HF+HF & 5.13 & 7.9$^a$ & D \\
        & TiO$_2$(R)& 1.86 & 1.89 & PBE+HF & 1.93 & 3.06$^e$ & D \\
        & TiO$_2$(A)& 1.97 & 2.14 & PBE+HF & 2.18 & 3.20$^e$ & I \\
        & {\bf MnO} & 1.67& 0.91 & HF+HF & 1.67 & 3.6$\sim$4.1$^{f}$ & I \\
        & {\bf In$_2$O$_3$}&1.87 & 1.16 & HF+HF & 1.87 & $\underline{<}$2.9$^{g}$ & I \\
 \hline
 \end{tabular}

 $^a$Ref.\cite{Madelung}, $^b$Ref.\cite{JAP94p3675}, $^c$Ref.\cite{ZnO-ZB1,ZnO-ZB2}, $^d$Ref.\cite{PRB39p10935}, $^e$Ref.\cite{Nanotechnology}, $^f$Ref.\cite{PRB44p1530,Huffman,Iskenderov,Kurmaev}, $^g$Ref.\cite{JAP37p299,JAP88p5180,PRB75p153205,PRL100p167402}
\end{table}

\vspace{0.3cm}
\textbf{3. Results and discussion}
\vspace{0.3cm}

Table I shows the calculated band gap using PBE, HF, and optimal PPs (Note that the optimal PP corresponds to the PP combination between PBE and HF which yields the largest band gap as will be further illustrated below.), in comparison with experiments. More details on the results can be found in Table S2 in the Supplemental Material. Compared to PBE PP, the HF PP yields a larger or at least comparable band gap in most cases. In other words, HF PPs produce results that are in better agreement with experiment. Such a favorable trend for the HF PPs is irrespective of the crystal structures. The only exceptions are the phosphorus compounds (i.e., GaP and InP). However, they are originated from different reasons. For GaP, the HF PP places the energy at $X$ point slightly below that at $\Gamma$ point, while for InP, the energy lowering directly occurs at $\Gamma$ point (see Fig. S1 in the Supplemental Material). Table I reveals noticeable gap opening for late-TM oxides, e.g., MnO, ZnO, CdO, and In$_2$O$_3$. By contrast, the band gaps between PBE and HF PPs just differ slightly for early-TM and alkaline oxides (TiO$_2$ and MgO).

%fig01
\begin{figure}[tbp]
\includegraphics[width=0.8\columnwidth]{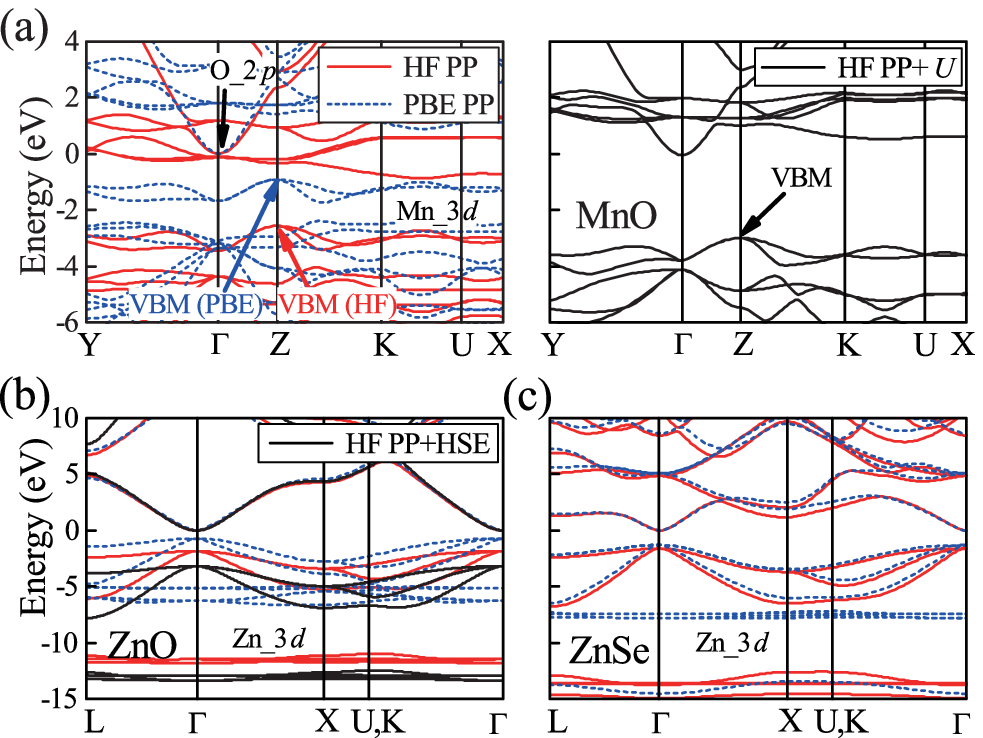}
\caption{\label{fig:fig1} (Color online) Band structures calculated by using HF (red solid lines) and PBE (blue dashed lines) PPs for (a) MnO, (b) zinc-blende ZnO, and (c) ZnSe. We take the lowest unoccupied non-$d$ states at $\Gamma$ which are anion $p$ states (mixed with the cation $s$ states in the cases of ZnO and ZnS) as the energy reference zero.
Right panel in (a) shows the DFT+$U$ band for MnO with $U =$ 4 eV. For ZnO in (b), besides the PBE and HF PP results, HSE results with HF PPs (black solid lines) are also shown. The high symmetric paths of the first Brillouin-Zone are shown in Fig. S2 of Supplemental Material.}
\end{figure}

Figure 1 shows the band structures of rocksalt MnO, zinc-blende ZnO, and zinc-blende ZnSe to illustrate the effect of the HF PPs on them. Results for other systems can be found in Fig. S1 of the Supplemental Material. Remarkably, the HF PPs change the band structure of MnO considerably, as can be seen in the left panel of Fig. 1(a) where a lowering of the Mn $d$-states happens on both the occupied and empty $d$-states but not much on their mutual repulsion. This leads to a considerably-lower empty $d$ states at the conduction band minimum. A lowering of the occupied $d$-states reduces the $p$-$d$ repulsion to the occupied O $p$-states, so the valence band maximum (VBM) in the HF PP calculation is also lowered. The net result is an opening of the band gap from that of PBE (0.91 eV) by 0.76 eV to 1.67 eV. The changes in the energy dispersion of the upper occupied bands resemble those by high-level self-interaction corrected DFT \cite{JPCM2p3973} and exact-exchange LDA/RPA calculations \cite{PRL103p036404, JPCM10p9241}, as can be more clearly seen in the right panel of Fig. 1(a) for which more discussion will be given later. The fact that the improvement due to HF PP on empty states is limited is also seen in other TM oxides, e.g., by the almost unchanged band structure of TiO$_2$ (See Fig. S1 in the Supplemental Material). It happens that the effect of the HF PP on magnetic properties is insignificant so in either calculations, MnO takes a type-II antiferromagnetic structure \cite{PRB64p024403} and the local magnetic moment on Mn is very much unchanged. As will be shown below, the deficiencies of the HF PP here can be offset by introducing a physics-based effective Hubbard $U$ on the semicore $d$-orbitals.

For zinc-blende ZnO in Fig. 1(b), we see that, due to the use of HF PPs, the Zn $3d$ states move $\sim$5 eV down to a position $\sim$10 eV below the VBM. Now the Zn $3d$ semicore gets much-less interactions with the O $p$ bands so its band width is reduced to only 1.1 eV. The highest-occupied oxygen $p$ band, on the other hand, moves down nearly rigidly with respect to that using PBE PPs to open up the band gap. ZnO is notorious for its too small calculated band gap. Therefore, it is useful to compare the result here with those calculated by other approaches: our standard PBE yields a gap of 0.68 eV, which is consistent with the previous result of 0.65 eV \cite{PRB09p235119}. The gap is increased to around 1.5 eV in Hubbard $U$ calculation \cite{Lany} or by using Engel-Vosko scheme \cite{Engel,pssb}. HSE hybrid functional calculation, on the other hand, yields a gap of 2.3 eV \cite{Lany}. Although $GW$ calculation \cite{Lany} can produce a gap of $\sim$3.3 eV that is in better agreement with experiment, this method itself is still a matter of controversy and it demands huge computational resources\cite{Liao}. By contrast, our HF PPs yield a gap of $\sim$1.9 eV, which is already comparable to that of HSE, yet the computational efficiency remains at that of DFT. Similar results are also found in wurtzite ZnO whose gap is increased from 0.86 eV of PBE to 2.13 eV of HF, as shown in Table I. For rocksalt CdO, which has a $4d$, instead of a $3d$, semicore, PBE PPs inaccurately predict a metallic behavior \cite{Burbano}, whereas HF PPs move down the Cd $4d$ bands by about 3 eV and the topmost valence bands to open an indirect gap of 0.69 eV (See Fig. S1 of the Supplemental Material).

%fig02
\begin{figure}[tbp]
\includegraphics[width=0.8\columnwidth]{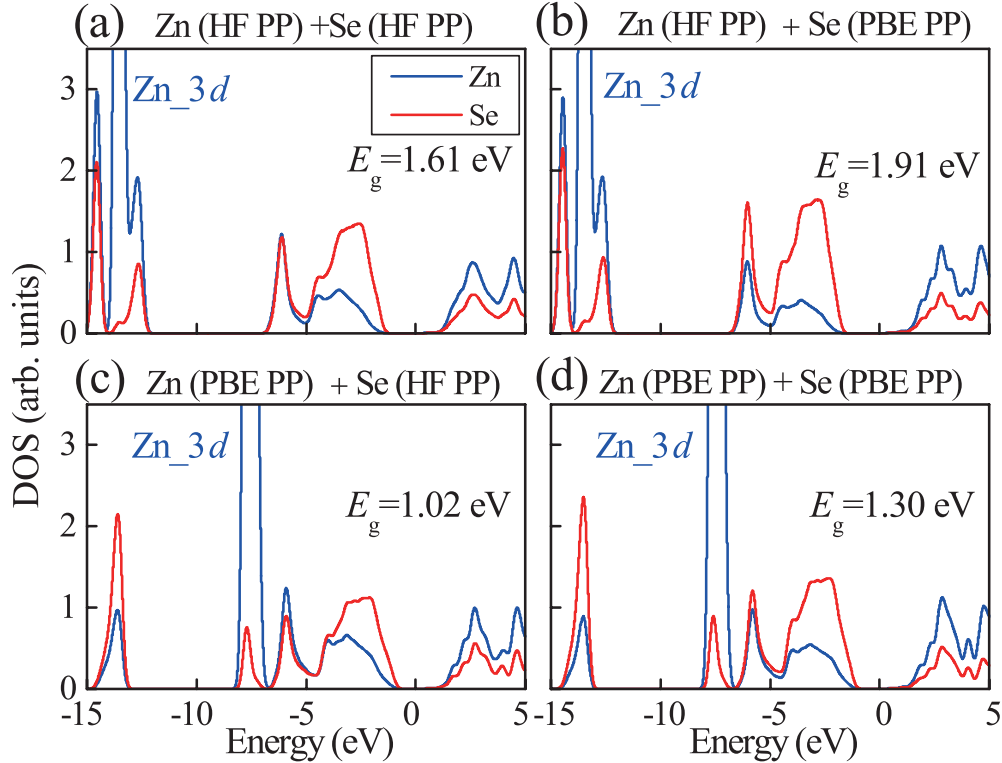}
\caption{\label{fig:fig2} (Color online) Partial densities of states of ZnSe obtained with the use of various combinations of  PPs: (a) Zn(HF)+Se(HF), (b) Zn(HF)+Se(PBE), (c) Zn(PBE)+Se(HF), and (d) Zn(PBE)+Se(PBE). The conduction band minimum is set as energy zero.}
\end{figure}

Se is the element in the same column as O in the Periodic Table.
Compared to ZnO, ZnSe in Fig. 1(c) has deeper Zn $3d$ states, as well as a smaller band gap due perhaps to the lower Se anti-bonding $p$ states (see Fig. 2). As a result of the deeper Zn $d$ states, the PBE PP band-gap error in ZnSe of 1.55 eV is smaller than that in ZnO of 2.59 eV. The improvement by HF PP follows the same trend, namely, 0.34 eV for ZnSe but 1.17 eV for ZnO. From a similar point of view, one can easily understand why DFT generally performs better for GaN and GaAs than for ZnO and ZnSe, say, the $3d$ semicore states of Ga lie much deeper in energy than those of Zn.

From our discussion in the beginning, in principle, the PPs can be either HF-type or PBE-type. To optimize the band gap, therefore, we use various combinations of PPs in the calculations and the optimal band gaps are summarized in Table I, as well as in Table S2 of the Supplemental Material. Figure 2 takes the ZnSe above as an example to illustrate the effect of the mixed PPs in some details, where the partial densities of states (PDOS) are plotted. Note that there are four possible combinations of the PPs, namely, (HF or PBE PP for Zn)$\otimes$(HF or PBE PP for Se), so we have four sub panels in Fig. 2. For simplicity, we may ignore the rich details in the PDOS but focus only on the band gap and energy positions of the Zn $3d$ semicore states. We see that both the band gap and the semicore position are determined by the type of Zn PP, insensitive to the type of Se PP. In other words, if a Zn HF PP is used, the system has a relatively larger gap of 1.61 and 1.91 eV, respectively, with a low-lying Zn $3d$ semicore at around $13$ eV below the VBM. If, on the other hand, a Zn PBE PP is used, the band gap of the system decreases by $\sim$0.6 eV with the Zn $3d$ semicore at only $6$ eV below the VBM.

%fig03
\begin{figure}[tbp]
\includegraphics[width=0.8\columnwidth]{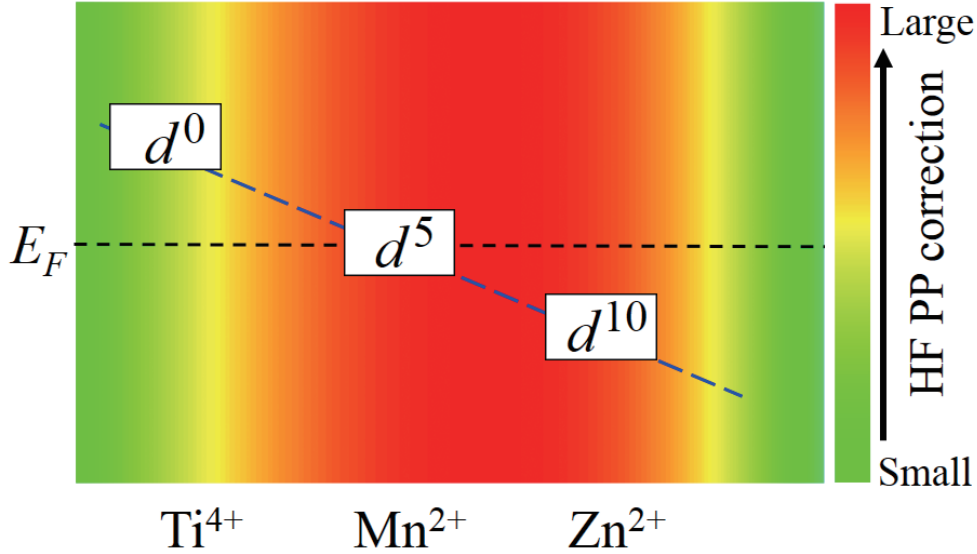}
\caption{\label{fig:fig3} (Color online) A schematic illustration of the expected correction to the band gap by HF PPs in systems containing semicore $d$ electrons. The energy positions of the $d$ states of different systems relative to the Fermi level ($E_F$ ) are also shown. The color represents the correction to the band gap by HF PPs. The higher the $d$-electron density and/or the closer to the $E_F$, the larger the correction.}
\end{figure}

Then, let us briefly discuss the distinct effects of HF PPs on different elements in a binary semiconductor as reflected by Table I. Often, self-interaction error is significant for the cation because losing valence electrons to the anion exposes its semicore $d$-electrons. Therefore, one may expect a HF PP should work better than a LDA/GGA PP. By contrast, the anion typically has much deeper core electrons while its valence electrons are more delocalized so the correlation effect is important but not the self-interaction error. Accordingly, the LDA/GGA PP is expected to be a better choice. For different cations involving $d$ semicores, the degree to which the HF PP can improve the band gap may also be different. As it turns out, there are two important factors, i.e., i) the electron density and ii) the energy position with respect to the Fermi level ($E_F$) of the $d$ electrons that determine the HF PP band-gap correction. Generally speaking, the higher the density and/or the closer to the $E_F$, the larger the correction. This is schematically illustrated in Fig. 3, as an example, for 3$d$ transition metal (TM) elements. For elements on the left of the figure such as Ti$^{4+}$, because its $d$ states are totally unoccupied, e.g., in TiO$_2$, little improvement can be expected from a HF PP. For element in the middle such as Mn$^{2+}$, on the other hand, because its $d$ states are half-occupied by five $d$ electrons across the Fermi level, e.g., in MnO, the influence of the HF PP on the band gap can be remarkable (here, for simplicity, we ignore the effect of spin polarization as it will not qualitatively change our physical picture). For elements on the right such as Zn$^{2+}$, while the $d$ states become fully occupied, their energies are moved away and down from $E_F$, e.g., in ZnO, thus it will be the interplay between the two factors mentioned above that determines the usefulness of the HF PP. Putting all together, the qualitative discussions here establish a simple yet remarkable trend in band-gap improvement that one may expect from using the HF PPs for TM compounds, which has been an insurmountable obstacle for LDA/GGA.

%fig04
\begin{figure}[tbp]
\includegraphics[width=0.8\columnwidth]{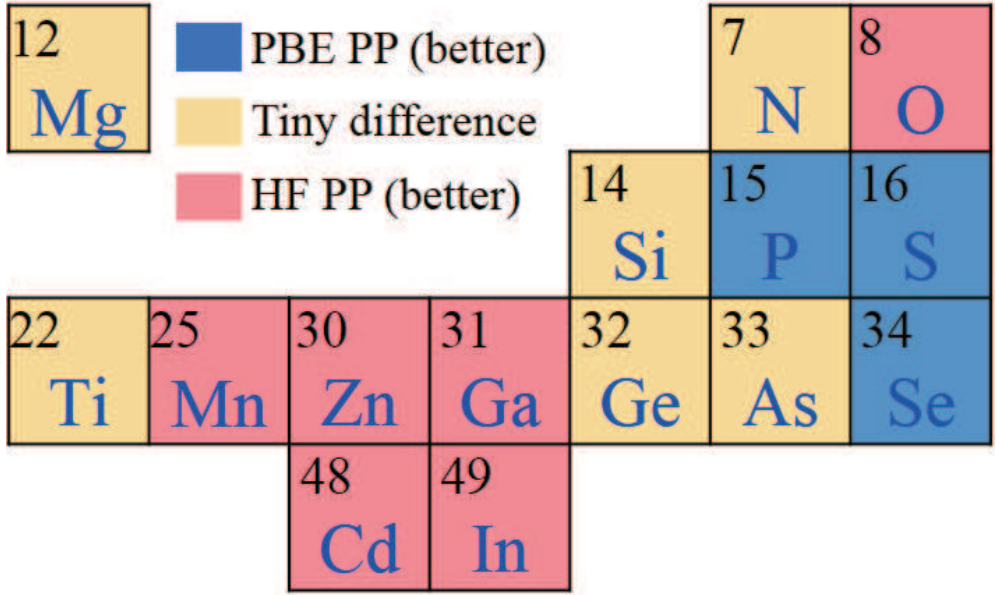}
\caption{\label{fig:fig4} (Color online) A map indicating the element-specific PP type, between HF PP and PBE PP, in the Periodic Table. The ``Tiny difference" means that the HF PP and PBE PP produce band gaps with very small difference.}
\end{figure}

The results in Fig. 2 reveal that the largest (optimal) band gap of 1.91 eV for ZnSe is obtained by a combination of Zn HF PP and Se PBE PP, rather than by using pure HF PPs or pure PBE PPs. It reinforces the notion that cation and anion may prefer a different type of PPs, as mentioned earlier. In Fig. 4, we summarize the element-specific PP type according to the calculated optimal band gaps in Table S2 of the Supplemental Material. It suggests that the HF PP works better for elements located in the central region of the Periodic Table, consisting mostly of the TMs, while the PBE PP works better for elements located in the lower right corner of the Table. For elements on the borderlines between the two, either PP produces similar results. Oxygen, due to its highly localized $p$ states \cite{PRB10p085207}, appears to be a special case and is also in favor of the HF PP. We have to emphasize that the optimal choice of PP must be sensitive to the element valence state, rather than fully determined by the element type, in particular for the early-TMs as the different valence states unambiguously correspond to different $d$-electron configurations.

The usefulness of HF PP can be further demonstrated through its ability to improve higher-level calculations, such as DFT+$U$ and HSE. For example for MnO, Fig. 1(a) shows that the insufficient spin splitting between the occupied and unoccupied $d$ states causes a too small band gap. This drawback can be fixed by DFT+$U$ method (As shown in the right panel of Fig. 1(a)). Using a $U$ = 4 eV together with HF PPs yields a band gap of 3.0 eV, which is on par with HSE result \cite{JPCC116p9876} but without extra computational cost. If we perform the HSE calculation alongside with the HF PPs, on the other hand, we obtain a band gap of 4.1 eV, which is in very good agreement with experiment \cite{PRB44p1530,Huffman,Iskenderov,Kurmaev}. Likewise, for ZnO, the HSE alongside with HF PPs yields a band gap of 3.2 eV, as shown in Fig. 1(b), which is also in very good agreement with experiment \cite{ZnO-ZB1,ZnO-ZB2}. The fact that we can get good results with no need of tuning the mixing parameters suggests that the HF PP may capture part of the essential physics of localized electrons.

\vspace{0.3cm}
\textbf{4. Conclusions}
\vspace{0.3cm}

In conclusion, we introduce a Hartree-Fock pseudopotential method, alongside with the standard DFT such as PBE for valence electrons, to perform electronic structure calculations. We show a systematic improvement of the band gap over a range of semiconductors of noticeably different physical properties, especially for those with occupied $d$ states. The method can be further improved by adopting the mixed PP approach. As the HF PPs are fully compatible with standard DFT calculations, no additional computational cost is required. While the DFT+$U$ method holds the same advantage, the system-dependent parameter $U$ not only deviates from the principles of {\sl ab initio} methods but also its application to atomic $s$ and $p$ states \cite{Tahini} is controversial. Last but not least, the HF PPs alongside with +$U$ or HSE yield significantly improved band gap for special systems such as MnO and ZnO. It is reasonable to speculate that the HF PPs may also be helpful for the GW calculations, such as the notorious band-gap convergence problem in ZnO.

\vspace{0.3cm}
\textbf{Acknowledgments}
%\vspace{-1.5em}
\vspace{0.3cm}

We thank Andrew M. Rappe and Yong Xu for helpful discussions. Work in China was supported by the Ministry of Science and Technology of China (Grant No. 2016YFA0301001), the National Natural Science Foundation of China (Grant Nos. 11674071, 11674188, and 11334006). SBZ was supported by the US Department of Energy (DOE) under Grant No. DESC0002623. Our computation is completed on the Explorer 100 cluster system of Tsinghua National Laboratory for Information Science and Technology.

\makeatletter
\def\bib@device#1#2{}
\makeatother
\makeatletter \renewcommand\@biblabel[1]{${#1}.$}\makeatother


\vspace{0.3cm}
\textbf{References}
\vspace{-1.5em}
\begin{thebibliography}{49}%
%\makeatletter
\bibitem{PNAS2017} J. P. Perdew, W. Yang, K. Burke, Z. Yang, E. K. U. Gross, M. Scheffler, G. E. Scuseria, T. M. Henderson, I. Y. Zhang, A. Ruzsinszky, H. Peng, J. Sun, E. Trushinm, A. G\"orling, \emph{Proc. Natl. Acad. Sci. USA}, 2017, \textbf{114}, 2801-2806.

\bibitem{Medvedev} M. G. Medvedev, I. S. Bushmarinov, J. W. Sun, J. P. Perdew, K. A. Lyssenko, \emph{Science}, 2017, \textbf{355}, 49-52.

\bibitem{Madelung} O. Madelung, $Semiconductors$: $Data$ $Handbook$ (Springer-Verlag, Berlin Heidelberg, 3rd edition, 2004).

\bibitem{Burbano} M. Burbano, D. O. Scanlon, G. W. Watson, \emph{J. Am. Chem. Soc.}, 2011, \textbf{133}, 15065-15072.

\bibitem{Toroker} M. C. Toroker, D. K. Kanan, N. Alidoust, L. Y. Isseroff, P. L. Liao, E. A. Carter, \emph{Phys. Chem. Chem. Phys.}, 2011, \textbf{13,} 16644-16654.

\bibitem{SBZPRB63} S. B. Zhang, S.-H. Wei, A. Zunger, \emph{Phys. Rev. B}, 2001, \textbf{63}, 075205.

\bibitem{Janotti} A. Janotti, C. G Van de Walle, \emph{Rep. Prog. Phys.}, 2009, \textbf{72}, 126501.

\bibitem{Baraff} G. A. Baraff, M. Schl\"{u}ter, \emph{Phys. Rev. B}, 1984, \textbf{30}, 3460.

\bibitem{Anisimov} V. I. Anisimov, J. Zaanen, O. K. Andersen, \emph{Phys. Rev. B}, 1991, \textbf{44}, 943.

\bibitem{Aulbur} W. G. Aulbur, L. J\"{o}nsson, J. W. Wilkins, \emph{Solid State Phys.}, 2000, \textbf{54}, 1-218.

\bibitem{SBZJPCM} S. B. Zhang, \emph{J. Phys.: Condens. Matter}, 2002, \textbf{14}, R881-R903.

\bibitem{HSE} J. Heyd, G. E. Scuseria, M. Ernzerhof, \emph{J. Chem. Phys.}, 2003, \textbf{118}, 8207-8215.

\bibitem{YangHY} J. Yang, L. Z. Tan, A. M. Rappe, \emph{arXiv:1707.04501}, 2017.

\bibitem{Chong} D. P. Chong, O. V. Gritsenko, E. J. Baerends, \emph{J. Chem. Phys.}, 2002, \textbf{116}, 1760-1772.

\bibitem{kollmar2007} C. Kollmara, M. Filatov, \emph{J. Chem. Phys.}, 2007, \textbf{127}, 114104-1-10.

\bibitem{Saidi} W. A. Al-Saidi, E. J. Walter, A. M. Rappe, \emph{Phys. Rev. B}, 2008, \textbf{77}, 075112.

\bibitem{OPIUM} http://opium.sourceforge.net.

%\bibitem{SM} See the Supplemental Material for more details. %The Hartree-Fock (HF) PP and the Hybrid (HY) PBE0 PP are firstly implemented in the newest version of this software. This makes it possible to generate two new types of PP for density functional theory calculations.

\bibitem{QE} P. Giannozzi, S. Baroni, N. Bonini, M. Calandra, R. Car, C. Cavazzoni, D. Ceresoli, G. L. Chiarotti, M. Cococcioni, I. Dabo, A. Dal Corso, S. de Gironcoli, S. Fabris, G. Fratesi, R. Gebauer, U. Gerstmann, C. Gougoussis, A. Kokalj, M. Lazzeri, L. Martin-Samos, N. Marzari, F. Mauri, R. Mazzarello, S. Paolini, A. Pasquarello, L. Paulatto, C. Sbraccia, S. Scandolo, G. Sclauzero, A. P. Seitsonen, A. Smogunov, P. Umari, R. M. Wentzcovitch, \emph{J. Phys.: Condens. Matter}, 2009, \textbf{21}, 395502.

\bibitem{PBE} J. P. Perdew, K. Burke, M. Ernzerhof, \emph{Phys. Rev. Lett.}, 1996, \textbf{77}, 3865.

\bibitem{JAP94p3675} I. Vurgaftman, J. R. Meyer, \emph{J. Appl. Phys.}, 2003, \textbf{94}, 3675-3696.

\bibitem{ZnO-ZB1} A. B. M. A. Ashrafi, A. Ueta, A. Avramescu, H. Kumano, I. Suemune, Y.-W. Ok, T.-Y. Seong, \emph{Appl. Phys. Lett.}, 2000, \textbf{76}, 550-552.

\bibitem{ZnO-ZB2} S.-K. Kim, S.-Y. Jeong, C.-R. Cho, \emph{Appl. Phys. Lett.}, 2003, \textbf{82}, 562-564.

\bibitem{PRB39p10935} P. E. Lippens, M. Lannoo, \emph{Phys. Rev. B}, 1989, \textbf{39}, 10935.

\bibitem{Nanotechnology} D. Reyes-Coronado, G. Rodr\'iguez-Gattorno, M. E. Espinosa-Pesqueira, C. Cab, R. de Coss, G. Oskam, \emph{Nanotechnology}, 2008, \textbf{19}, 145605.

\bibitem{PRB44p1530} J. van Elp, R. H. Potze, H. Eskes, R. Berger, G. A. Sawatzky, \emph{Phys. Rev. B}, 1991, \textbf{44}, 1530.

\bibitem{Huffman} D. R. Huffman, R. L. Wild, M. Shinmei, \emph{J. Chem. Phys.}, 1969, \textbf{50}, 4092-4094.

\bibitem{Iskenderov} R. N. Iskenderov, I. A. Drabkin, L. T. Emel'yanova, Y. M. Ksendzov, \emph{Fiz. Tverd. Tela}, 1968, \textbf{10}, 2573.

\bibitem{Kurmaev} E. Z. Kurmaev, R. G. Wilks, A. Moewes, L. D. Finkelstein, S. N. Shamin, J. Kune{\v{s}}, \emph{Phys. Rev. B}, 2008, \textbf{77}, 165127.

\bibitem{JAP37p299} R. L. Weiher, R. P. Ley, \emph{J. Appl. Phys.}, 1966, \textbf{37}, 299-302.

\bibitem{JAP88p5180} V. Christou, M. Etchells, O. Renault, P. J. Dobson, O. V. Salata, G. Beamson, R. G. Egdell, \emph{J. Appl. Phys.}, 2000, \textbf{88}, 5180-5187.

\bibitem{PRB75p153205} P. Erhart, A. Klein, R. G. Egdell, K. Albe, \emph{Phys. Rev. B}, 2007, \textbf{75}, 153205.

\bibitem{PRL100p167402} A. Walsh, J. L. F. Da Silva, S.-H. Wei, C. Korber, A. Klein, L. F. J. Piper, A. DeMasi, K. E. Smith, G. Panaccione, P. Torelli, D. J. Payne, A. Bourlange, R. G. Egdell, \emph{Phys. Rev. Lett.}, 2008, \textbf{100}, 167402.

\bibitem{JPCM2p3973} V I Anisimov, M A Korotin, E Z Kurmaev, \emph{J. Phys.: Condens. Matter}, 1990, \textbf{2}, 3973-3987.

\bibitem{PRL103p036404} E. Engel, R. N. Schmid, \emph{Phys. Rev. Lett.}, 2009, \textbf{103}, 036404.

\bibitem{JPCM10p9241} T. Kotani, \emph{J. Phys.: Condens. Matter}, 1998, \textbf{10}, 9241-9261.

\bibitem{PRB64p024403} J. E. Pask, D. J. Singh, I. I. Mazin, C. S. Hellberg, J. Kortus, \emph{Phys. Rev. B}, 2001, \textbf{64}, 024403.

\bibitem{PRB09p235119} M. Topsakal, S. Cahangirov, E. Bekaroglu, S. Ciraci, \emph{Phys. Rev. B}, 2009, \textbf{80}, 235119.

\bibitem{Lany} S. Lany, A. Zunger, \emph{Phys. Rev. B}, 2010, \textbf{81}, 113201.

\bibitem{Engel} E. Engel, S. H. Vosko, \emph{Phys. Rev. B}, 1993, \textbf{47}, 13164.

\bibitem{pssb} Z. Charifi, H. Baaziz, A. H. Reshak, \emph{Phys. Stat. Sol. B}, 2007, \textbf{244}, 3154-3167.

\bibitem{Liao} P. L. Liao, E. A. Carter, \emph{Phys. Chem. Chem. Phys.}, 2011, \textbf{13,} 15189-15199.

\bibitem{PRB10p085207} P. Dev, P. Zhang, \emph{Phys. Rev. B}, 2010, \textbf{81}, 085207.

\bibitem{JPCC116p9876} D. K. Kanan, E. A. Carter, \emph{J. Phys. Chem. C}, 2012, \textbf{116}, 9876-9887.

\bibitem{Tahini} H. Tahini, A. Chroneos, R. W. Grimes, U. Schwingenschlogl, A. Dimoulas, \emph{J. Phys.: Condens. Matter}, 2012, \textbf{24}, 195802.

\end{thebibliography}
\end{document}